\documentclass[aps,twocolumn,groupedaddress,superscriptaddress,amsmath,amssymb,prl]{revtex4-2}

\usepackage[utf8x]{inputenc}
\usepackage{ucs}
\usepackage{graphicx}
\usepackage{amsfonts}
\usepackage{amsmath}
\usepackage{float}
\usepackage{physics}
\usepackage{mathtools}
\usepackage{bbold}
\usepackage{amsthm}
\usepackage[breaklinks=true,%
            colorlinks=true,%
            linkcolor=blue,%
            urlcolor=blue,%
            citecolor=blue]{hyperref}

\usepackage{color}

\usepackage{xr}
\makeatletter
\newcommand*{\addFileDependency}[1]{% argument=file name and extension
  \typeout{(#1)}
  \@addtofilelist{#1}
  \IfFileExists{#1}{}{\typeout{No file #1.}}
}
\makeatother

\newcommand*{\myexternaldocument}[1]{%
    \externaldocument{#1}%
    \addFileDependency{#1.tex}%
    \addFileDependency{#1.aux}%
}

\myexternaldocument{supp}
\newcommand{\av}[1]{\mathbb{E}\left(#1\right)}
\newcommand{\ndim}{N}

\makeatletter
\renewcommand*\env@matrix[1][*\c@MaxMatrixCols c]{%
  \hskip -\arraycolsep
  \let\@ifnextchar\new@ifnextchar
  \array{#1}}
\makeatother
\def\env@matrix{\hskip -\arraycolsep
  \let\@ifnextchar\new@ifnextchar
  \array{*\c@MaxMatrixCols c}}

\begin{document}

%\title{Classically optimized QAOA algorithm for hard integer value problems}

\title{Quantum mean value approximator for hard integer value problems}

\author{David Joseph}
\affiliation{Electrical and Electronic Engineering Department, Imperial College London}%
\affiliation{Physics Department, Imperial College London}

\author{Antonio J. Martinez}
\affiliation{Sandbox@Alphabet}%

 \author{Cong Ling}
 \affiliation{Electrical and Electronic Engineering Department, Imperial College London}%
 
 \author{Florian Mintert}
 \affiliation{Physics Department, Imperial College London}%

\begin{abstract}
Evaluating the expectation of a quantum circuit is a classically difficult problem known as the quantum mean value problem (QMV). It is used to optimize the quantum approximate optimization algorithm and other variational quantum eigensolvers. We show that such an optimization can be improved substantially by using an approximation rather than the exact expectation. Together with efficient classical sampling algorithms, a quantum algorithm with minimal gate count can thus improve the efficiency of general integer-value problems, such as the shortest vector problem (SVP) investigated in this work.
\end{abstract}

\date{\today}%10 Feb 2021}

\maketitle

Quantum computers have the potential to solve certain computational problems which remain intractable on classical hardware, such as factorizing integers \cite{Shor1999Polynomial-timeComputer}, which breaks the security of most presently used public-key cryptosystems \cite{Rivest1978ACryptosystems, Diffie1976NewCryptography}. Such quantum devices are extremely fragile however, meaning that in the near term only those algorithms which apply minimal quantum gates will be practicable on Noisy Intermediate-Scale Quantum (NISQ) computers \cite{Preskill2018QuantumBeyond, Bharti2021NoisyAlgorithms}. The quantum approximate optimization algorithm (QAOA) \cite{Farhi2014AAlgorithm} is one algorithm of interest \cite{Streif2019ComparisonAnnealing, Farhi2019TheSize}, due to its low depth variants, which are a coarse approximation to the continuous adiabatic quantum algorithm (AQA) \cite{Farhi2000QuantumEvolution, Albash2018AdiabaticComputation}.
In contrast to AQA that is based on the adiabatic transition between a simple Hamiltonian and the problem Hamiltonian whose ground state contains the solution of the problem,
QAOA is based on a sequence of gates induced by these two Hamiltonians. 
While AQA  is by definition slow and so requires hardware with long coherence times, the length of the gate sequence in QAOA can be reduced to a minimum, making it compatible with limited coherence times in NISQ devices \cite{Yang2017OptimizingPrinciple}.
In the limit of short sequences, however, the success of QAOA is highly sensitive to the optimal choice of Rabi-angles in the constituent gates.

Identification of such optimal parameters, necessarily requires the ability to estimate the algorithm's performance for any value of the Rabi-angles \cite{Bittel2021TrainingSystems}.
The energy expectation value is a widely used figure of merit for scoring QAOA (and other variational quantum algorithms \cite{Cerezo2020VariationalAlgorithms}),
but its evaluation is also a classically hard problem.
Consequently, methods to approximate this value have been developed using tensor network contractions \cite{Markov2008SimulatingNetworks, Aaronson2016Complexity-theoreticExperiments} and other techniques \cite{Bravyi2021ClassicalValues}. Such methods are only applicable for systems which are locally connected, however, meaning there are no classically efficient ways to compute the quantum mean value (QMV) for fully connected systems, which are deemed to be of most practical use for the lowest depth variants of QAOA \cite{Farhi2020TheCase}.

One computational problem which is subject to intense scrutiny is the shortest vector problem (SVP) \cite{Ajtai1996GeneratingAbstract}, the hardness of which underpins the security of a family of cryptosystems expected to replace the provably quantum-insecure ones of today \cite{Peikert2016ACryptography, Regev2005OnCryptography, Goldreich1997Public-KeyProblems}.
Solving large instances of SVP (i.e. finding short vectors in high-dimensional lattices) is intractable using the best classical algorithms \cite{Gama2010LatticePruning, Nguyen2008SievePractical},
but practical solutions could be found in terms of sampling if sampling from suitable probability distributions were possible \cite{Aggarwal2015SolvingSampling, Gentry2008TrapdoorsConstructions}.
Solving SVP efficiently (using either quantum or classical techniques) would in effect condemn all lattice-based cryptosystems, which are a frontrunner in the US government's process to standardize quantum-resistant cryptosystems \cite{Alagic2019StatusProcess, Jeffrey1998NTRUCryptosystem, Bos2018CRYSTALS-Kyber:KEM}.

It is known that a QAOA algorithm of minimal length cannot be efficiently classically simulated~\cite{Farhi2016QuantumAlgorithm}, and while it cannot be expected to provide the actual solution with high probability, we will show that it can be used as a sampler that returns short vectors, which then can be employed as input to classical sampling algorithms.

The  extreme case of QAOA 
\begin{equation}
\label{eq:state}
\ket{\psi}=\exp(-i\beta H_D)\exp(-i\gamma H_P)\ket{\psi_0}\ ,
\end{equation}
with minimal gate count is given in terms of one single evolution induced by the problem Hamiltonian $H_P$ and one single evolution induced by a driver Hamiltonian $H_D$. 
The initial state $\ket{\psi_0}$ is typically taken to be the equal superposition of all feasible solutions \cite{Coppersmith2002AnFactoring},
and the state $\ket{\psi}$ should approximate the ground state $\ket{\psi_S}$ of $H_P$ that provides the solution to the problem.
While in the case of long gate sequences, suitable values for the Rabi angles can be determined from first principles,
there is no obvious choice for the parameters $\beta$ and $\gamma$ in the low depth versions of QAOA.
Since the extent to which $\ket{\psi}$ is a good approximation to $\ket{\psi_S}$ typically depends very sensitively on the parameter choices of $\beta$ and $\gamma$,
it is essential to determine a suitable parameter set before running the algorithm.
Even though the present problem does not require QAOA to identify the exact ground state of the problem Hamiltonian,
the efficiency of a sampling algorithm will be higher if states generated with QAOA are closer to the ground-state.
It is thus advisable to use the parameter values for $\beta$ and $\gamma$ that minimize the expectation value
\begin{equation}
    \mu = \bra{\psi}  H_P \ket{ \psi}
\label{eq:mu}
\end{equation}
of the problem Hamiltonian with respect to the state $\ket{\psi}$.

To be specific, the following discussion will be focussed on Hamiltonians of the form 
\begin{equation}
\label{eq:simple_ham}
    H_P = \sum_{ij=1}^{\ndim}  \mathbf{G_{ij}} \hat{Q}_i \hat{Q}_j \ ,
\end{equation}
with a set of mutually commuting operators $\hat{Q}_i$, each of which has only integer eigenvalues,
and a coefficient matrix ${\bf G}$ encoding the problem to be solved.
This type of Hamiltonian is, in particular, suitable for problems based on a lattice,
{\it i.e.} a repeating pattern of points in $\ndim$-dimensional space, described by $\ndim$ linearly independent basis vectors $\textbf{b}_i$.
The goal of SVP is the identification of the non-zero vector $\textbf{v} = \sum_{i=1}^{\ndim}x_i \textbf{b}_i$ with minimal norm $\| \textbf{v} \|$ and integer expansion coefficients $x_i$.
With the coefficient matrix ${\bf G}$ in Eq.~\eqref{eq:simple_ham} taken to be the Gram matrix with elements $\mathbf{G_{ij}}=\textbf{b}_i\textbf{b}_j$ containing the overlap of the lattice basis vectors, the eigenvalues of $H_P$ are the squared lengths $\| \textbf{v} \|^2$ of all the different lattice vectors \cite{Joseph2020Not-so-adiabaticProblem, Joseph2021TwoProblem}, with the ground state corresponding to the zero vector, the first excited state corresponding to the shortest nonzero lattice vector, and so forth.
The problem is then the quest for the first excited state of $H_P$.
Since this problem Hamiltonian  is fully connected, previous works on QMV that utilize tensor network contractions \cite{Markov2008SimulatingNetworks, Aaronson2016Complexity-theoreticExperiments}, or require local Hamiltonians \cite{Bravyi2021ClassicalValues} are not applicable.

For a practical implementation, the operators $\hat{Q}_i$ need to be expressed in terms of qubit operators.
In order to realise an integer spectrum in the range $[-2^k, 2^k - 1]$ for each of the operators $\hat{Q}_i$, $k$ qubits per operator $\hat{Q}_i$ are required.
A natural choice is given by
\begin{equation}
\label{eq:qudit}
    \hat{Q}_i = \sum_{p=0}^k 2^{p-1} Z_{ip} + \frac{\mathbb{1}}{2}\ ,
\end{equation}
where the operator $Z_{ip}$ denotes the Pauli Z operator for qubit $p$ in the set of qubits realising $\hat{Q}_i $.

With this explicit encoding, the expectation value $\mu$ (Eq.\eqref{eq:mu}) to be minimised is given by
\begin{equation}
\label{eq:anal_exp}
    \mu = \frac{1}{4} \sum_{ij}^N \mathbf{G_{ij}} \Bigg[ 1 + \sum_{p,q}^k 2^{p+q} \Gamma_{ijpq}
    + \sum_{p=0}^k 2^p \Big( \Omega_{ip} + \Omega_{jp} \Big) \Bigg]\ ,
\end{equation}
in terms of the expectation values $\Gamma_{ijpq}=\bra{\psi}Z_{ip} Z_{jq}\ket{\psi}$, and
$\Omega_{ip}=\bra{\psi}Z_{ip}\ket{\psi}$
of the two-body interaction and single-qubit energy terms in the Hamiltonian.
The explicit dependence on $\gamma$ is given in Supp. Mat. Eqs. \eqref{eq:omega_def}, \eqref{eq:gamma_def}.

\begin{figure}[t]
    \centering
    \includegraphics[width=\linewidth]{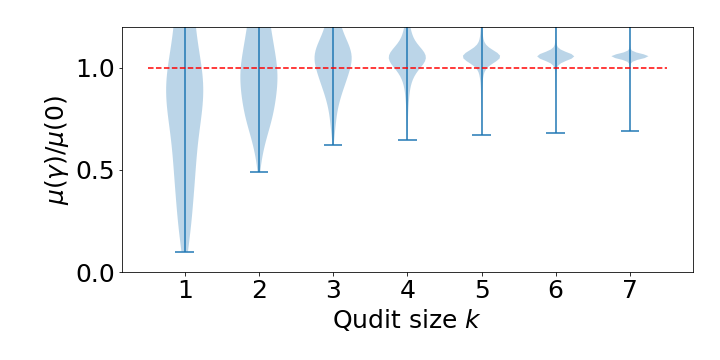}
    \caption{Each violin plot shows the distribution of $\mu (\gamma)/ \mu(0)$ over $\gamma$ obtained from $45$ two-dimensional lattices, numerically simulated using TensorFlow Quantum \cite{Broughton2020TensorflowLearning}.
    With growing system size $k$, the distributions get increasingly narrow, and the centers of the distributions lie above the classical baseline (red dashed line).
    The tail of the distributions, however extend substantially below this baseline, indicating clear improvement over classical sampling for rare values of $\gamma$.}
    \label{fig:violin}
\end{figure}
As an illustrative example, we will discuss the  QAOA circuit (the final state of which is given in Eq.\eqref{eq:state}) with local driver Hamiltonian $H_D = \sum_j X_j$ and fixed Rabi angle $\beta=\pi/4$ for $H_D$.
This reduces the variational parameter space to the single scalar parameter $\gamma$,
so that for a given problem instance, the expectation $\mu$ in Eq.~\eqref{eq:anal_exp} depends solely on this variable, {\it i.e.} $\mu=\mu(\gamma)$.

The violins in Fig.~\ref{fig:violin} depict the probability density of $\mu(\gamma)/\mu(0)$ for $\gamma$ uniformly distributed in the range $[0, \pi]$ for two-dimensional lattices and qudit size ranging from $k=1$ to $k=7$.
The importance of $\mu(0)$ is that it is equivalent to the uniform random lattice vector sampler, and so using $\mu(\gamma)/\mu(0)$ as a figure of merit enables one to compare the quantum algorithm with a classical baseline (red dashed line in Fig.~\ref{fig:violin}) and average across different lattices.
As one can see, the distributions become increasingly narrow with growing system size $k$.
Apart from the small-qudit cases $k=1$ and $k=2$, the centers of the violins lie above the base-line;
a low depth QAOA sampling algorithm will thus perform worse than the classical sampler for the vast majority of values of $\gamma$.
The lower end of the violins, however, indicate that there are values of $\gamma$ that result in substantially lower values of $\mu$, {\it i.e.} substantially improved performance in a sampling algorithm.
Identification of such an atypical value of $\gamma$ is necessary in order to benefit from such a low depth implementation of QAOA for an SVP sampling algorithm.

The identification of such a rare value of $\gamma$ requires the ability to evaluate Eq.~\eqref{eq:anal_exp}.
Since in any relevant application the construction of the desired state vectors following Eq.~\eqref{eq:state} is out of reach by classical means,
it seems that also any evaluation of $\mu$ has to be performed on a quantum-mechanical device.
This, however, is not the case, since the transformed operators
\begin{equation}
e^{i\gamma H_P}e^{i\beta H_D}Z_{ip}e^{-i\beta H_D}e^{-i\gamma H_P}
\end{equation}
involve only a fraction of the total number of qubits, and their expectation values with respect to the simple initial state $\ket{\psi_0}$ is analytically accessible.
The function $\mu(\gamma)$ to be minimised can therefore be evaluated classically even though the corresponding QAOA algorithm truly requires quantum hardware.

This classical evaluation can be facilitated substantially, noticing that 
the expectation $\mu$ can be approximated in terms of a small number of dominant terms, whose large magnitudes derive from the exponential prefactors of Eq. $\eqref{eq:qudit}$.
The expression
\begin{equation}
\label{eq:approx_exp}
    \mu \simeq \mu_A=\frac{1}{4} \sum_{ij}^N \mathbf{G_{ij}}  \sum_{k-A < p,q\le k} 2^{p+q} \Gamma_{ijpq}\ ,
\end{equation}
with $A\gtrsim 0$,
yields an accurate estimate for $\mu$ in terms of a small number of summands.
The value of $A$ defines the order of approximation in terms of the number of qubits that contribute to the operators $\hat Q_i$ in Eq.~\eqref{eq:qudit}.
The lowest order approximation $A=1$ is based entirely on the most significant qubits and the most accurate approximation $A=k$ takes into account all the qubits and only neglects the single-qubit contributions in Eq.~\eqref{eq:anal_exp}.
Since the computational effort for single-qubit contributions is negligible for large systems, the asymptotic cost of computing $\mu_A$ relative to $\mu$ is $O(A^2 / k^2)$.

\begin{figure}[t]
    \centering
    \includegraphics[width=\linewidth]{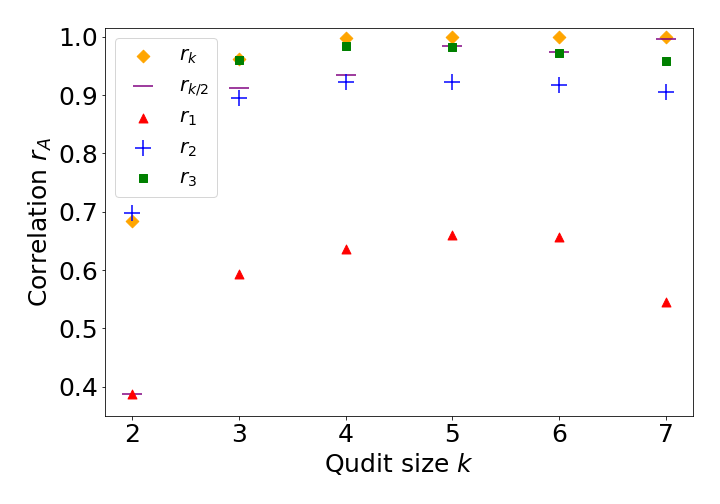}
    \caption{
    % \flo{font size axis labels}
    Pearson correlation coefficient $r_A$ for $A=1,2,3,k/2$ and $k$ averaged over $45$ two-dimensional lattices.
    Even approximations $\mu_A$ based on a few most-significant qubits, i.e. small values of $A$, result in an excellent approximation of the exact energy expectation value $\mu$ as indicated by a value of $r_A$ close to unity.
    %\flo{add $\mu_3$, drop random, maybe logscale for $1-r_A$.}
    }
    \label{fig:approx_correlations}
\end{figure}

The approximators $\mu_A$ can be compared to the exact expectation $\mu$ with the Pearson correlation coefficient
\begin{equation}
r_A = \frac{\mbox{cov}(\mu_A,\mu)}{\sigma_{\mu_A}\sigma_\mu}\ ,
\end{equation}
with the covariance $\mbox{cov}(x,y)=\av{xy}-\av{\av{x}\av{y}}$
and variance $\sigma_x^2=\mbox{cov}(x,x)$, where $\av{x}$ is a short-hand notation for the average of $x(\gamma)$ with $\gamma$ ranging from $0$ to $\pi$.

Fig~\ref{fig:approx_correlations} shows the averaged Pearson correlation coefficients $r_A$ for a range qudit sizes, with the average taken over $45$ different two-dimensional lattices. An approximation in terms of only the most significant qubits is rather rough, as $r_1$ does not exceed values of about $0.6$.
Taking into account the two most significant qubits, however, already results in a substantially better approximation with a Pearson correlation reaching $r_2\simeq 0.9$.
The approximation can be further improved by taking into account the next most significant qubits, and comparison of $r_{k/2}$ (purple dashes) and $r_k$ (yellow diamonds) demonstrates the rapid convergence, which thus 
guarantees a very good approximation in terms of a small fraction of all qubits.

\begin{figure}[t]
    \centering
    \includegraphics[width=\linewidth]{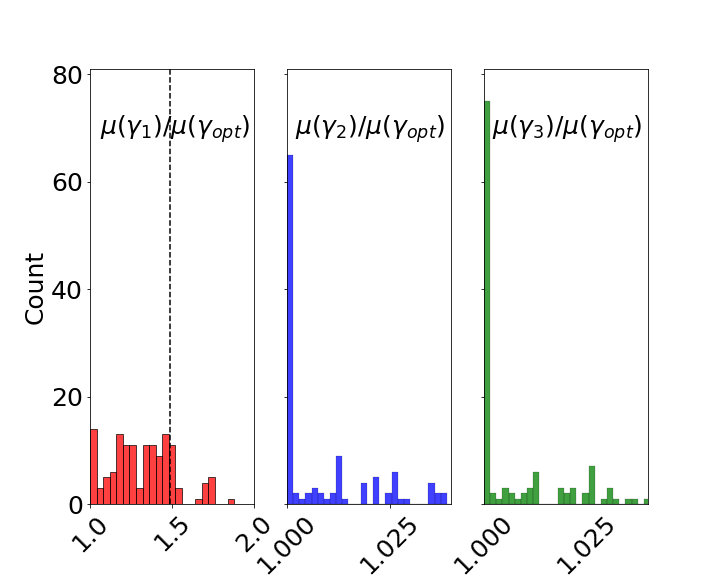}
    \caption{
    Histogram for the ratio $\mu(\gamma_A)/\mu(\gamma_{opt})$ where $\gamma_A$ and $\gamma_{opt}$ are the values of $\gamma$ that minimze $\mu_A$ and $\mu$ respectively. Each distribution contains datapoints for 45 lattices with $k=5,6,7$.
    The bin width is $0.04$ for $A=1$ and $1.5\ 10^{-3}$ for $A=2$ and $A=3$. The black dashed vertical in left hand plot denotes $\mu(0)/\mu(\gamma_{opt})$, but this value is too large to be shown for $A=2,3$.
   The actual minimum of $\mu$ is approximated increasingly well by $\gamma_A$ with increasing value of $A$, and even coarse approximations ($A\ll k$) yield excellent results.}
    \label{fig:minima}
\end{figure}
The high correlation $r_A$ for rather small values of $A$ suggest that minimisation of $\mu_A$ helps to identify a value of $\gamma$ that results in a low value of $\mu$, and consequently this value of $\gamma$ makes a QAOA algorithm well suited for an SVP sampler.
For an approximator $\mu_A$ to be useful, however, it must not just correlate well with the exact expectation $\mu$, it must also agree closely with $\mu$ on the minima of the parameter landscape, which in this case is a $1$-dimensional landscape over $\gamma$.
To demonstrate this, Fig.~\ref{fig:minima} depicts histograms for the ratio $\mu(\gamma_A)/\mu(\gamma_{opt})$,
where $\gamma_A$ is the value of $\gamma$ that achieves the minimum of $\mu_A$ and $\gamma_{opt}$ is the value of $\gamma$ that achieves the minimum of $\mu$, for $k=5,6,7$, obtained from $45$ two-dimensional lattices.
This figure of merit is insightful because it shows how well a $\gamma$ deduced from classical optimization of $\mu_A$ might perform when used on a fully quantum run of the algorithm. For a perfect approximator, one would see a single tall bar at $1.0$ indicating that results of optimization of the approximation are indistinguishable from optimization of the exact expectation.
From Fig.~\ref{fig:violin} the classical sampler, equivalent to $\mu(0)$, achieves an expectation of $\approx 1.5$ times higher than the minimal $\mu$, and this is shown in Fig. \ref{fig:minima} by the black dashed vertical in the $A=1$ histogram (though this value is far too high to be shown on the $A=2,3$ histograms). The fact that most of the density of the $A=1$ histogram is to the left of the black dashed line demonstrates that even using the coarsest approximation, one can identify $\gamma$ values that significantly outperform the classical sampler.
For $A=2$ and $A=3$, the deviation of $\mu(\gamma_A)/\mu(\gamma_{opt})$ from its ideal value of $1.0$ is at most a few percent, and in the vast majority of cases it is substantially smaller.
The reduction in performance due to optimization of an approximation of $\mu$ instead of the exact function is therefore substantially smaller than the improvement of an exactly-optimized QAOA algorithm over a classical sampler.

For the sake of concreteness, the optimization of low depth QAOA algorithms is discussed here in the context of the shortest vector problem, which is important in post-quantum cryptography.
The central ideas, however, apply to a great variety of quantum algorithms, where the problem Hamiltonians have widely ranging interaction magnitudes.
Integer programming problems \cite{Papadimitriou1998CombinatorialComplexity, Wolsey1999IntegerOptimization}, of which the knapsack problem is a simple example \cite{Mathews1896OnNumbers} fall into this category, since a binary representation of qudits is very natural in those cases.

Since unoptimized versions of QAOA with minimal circuit length fail to result in an algorithmic improvement, but optimized versions can clearly outperform what is classically achievable,
the present optimization techniques promise to bring minimal quantum algorithms closer to practical use.
Once existing hardware for quantum information processing gives access to longer gate sequences than the minimal version of QAOA considered here, the optimization problem naturally generalizes into an optimization over more parameters and a function to be optimized that  is computationally more expensive to evaluate than in the situation discussed here.
It is likely that even with approximations such functions are too costly to evaluate by classical means.
The approximations introduced here apply straight-forwardly also to the fully quantum version of the algorithm, as one can measure $\ket{\psi}$ with respect to $H_A$ , as per Eq. \eqref{eq:approx_ham_def} of Supp. Mat.. Similar to their reduction of effort in classical computation, this approach results in a reduced gate count in a quantum mechanical evaluation.
The techniques of the paper are thus not limited to the current era of NISQ devices, but they will find similar applicability in future generations of quantum mechanical devices.

\textit{Acknowledgements} The authors would like to thank Adam Callison for many discussions throughout the work, Stefan Leichenauer for his advice on approximators, and Guillaume Verdon  for insteresting conversations about qudit-based QAOA. The work was supported in part by EPSRC.

% -- {\bf discussion partners and funding}

\onecolumngrid
\appendix
\newpage

\setcounter{equation}{0}
\setcounter{figure}{0}
\setcounter{table}{0}
\setcounter{page}{1}
\makeatletter
\renewcommand{\theequation}{S\arabic{equation}}
\renewcommand{\thefigure}{S\arabic{figure}}
\renewcommand{\citenumfont}[1]{S#1}

\section{Supplementary Material}
% \maketitle

While the expectation $\mu$ and its approximators $\mu_A$ are defined in the main text, this material provides explicit analytical formulae for the dependence of $\Omega, \Gamma$ on the choice of parameter $\gamma$, as well as an alternate (but closely related) interpretation of $\mu_A$ in terms of approximate Hamiltonians, which provides insight into how the work can be applied in the fully quantum realm.

\subsection{Approximator Hamiltonian Definition}
\label{sec:approx_ham_def}
The approximators $\mu_A$ of Eq. \eqref{eq:approx_exp} can alternatively be expressed directly as the expectation of truncated Hamiltonians $H_A$. This is because the expectation approximators can be written $\mu_A = \bra{\psi} H_A \ket{\psi}$ where $H_A$ consists of the fully connected graph of two-qubit operators of $H_P$ between the $A$ most significant qubits
\begin{equation}
\label{eq:approx_ham_def}
H_A = \frac{1}{4} \sum_{ij} \mathbf{G_{ij}} \sum_{k-A < p,q \leq k}^k 2^{p+q} Z_{ip} Z_{jq}.
\end{equation}
It is important to note that in this representation of $\mu_A$, the expectation is taken with respect to an approximate Hamiltonian $H_A$, but the final state $\ket{\psi}$ is generated by the unitaries induced by the driver $H_D$ and the \textit{full} problem Hamiltonian $H_P$. To use $H_A$ in the synthesis of $\ket{\psi}$ would fundamentally oversimplify the underlying classical problem, and this was confirmed numerically by poor approximations.

\subsection{Expectation formulae}
\label{sec:exp_def}

Eq \eqref{eq:anal_exp} gives $\mu$ in terms of $\Omega, \Gamma$ which are the expectation of single and two-qubit operators of $H_P$. The expectations $\Omega, \Gamma$ are defined shortly after Eq \eqref{eq:anal_exp} in the main text, but this material provides the explicit dependencies of $\Omega, \Gamma$ on the variational parameter set, in this case $-\gamma$ (in order to reduce the number of minus signs in the following). Due to the common occurrence of such trigonometric terms, denote the shorthand
\begin{equation}
    \begin{split}
        \mathcal{S}_{ip} =  \sin(2^p \gamma \sum_v^N \mathbf{G_{iv}}) =  \sin(2^p \gamma \sum_v^N \mathbf{G_{vi}}),\\
        \mathcal{C}_{ip} =  \cos(2^p \gamma \sum_u^N \mathbf{G_{iv}}) = \cos(2^p \gamma \sum_u^N \mathbf{G_{vi}}).
    \end{split}
\end{equation}
For a given SVP instance, the problem Hamiltonian is determined by the Gram matrix $\mathbf{G}$ and the number of qubits per qudit $k$.

\subsubsection{Single-qubit operator expectation}
Then the expectation of the single-qubit operator is
\begin{equation}
\label{eq:omega_def}
    \Omega_{ip} = \frac{ -  {\cal S}_{ip}}{\cos(2^{2p} \gamma \mathbf{G_{ii}} ) }
    \prod_{v=1}^N \prod_{s=0}^k \cos(\mathbf{G_{iv}} 2^{p+s} \gamma )\ ,%{\cal S}_{p+s,i,v}\ ,
\end{equation}
where the indices $(i,p)$ describe the location of the qubit, with $i$ being the corresponding basis vector, and $p$ being the significance of the qubit within qudit $\hat{Q}_i$.

The expectation of $Z_{ip}$ with respect to the evolved state $\ket{\psi}$ is
\begin{equation}
\label{eq:single}
    \Omega_{ip} =  \bra{\psi_0} e^{-i \gamma H_P} e^{i \beta H_D} Z_{ip} e^{-i \beta H_D} e^{i \gamma H_P} \ket{\psi_0},
\end{equation}
where $\ket{\psi}$ is expressed in the form of Eq. \eqref{eq:state} (but with the substitution of $-\gamma$ for $\gamma$).
Conjugation of $Z_{ip}$ with the driver Hamiltonian and $\beta=\pi/4$ transforms $Z_{ip}$ to $Y_{ip}$.
After conjugating $Y_{ip}$ with the single-qubit part of $H_P$, the transformed operators inside of $\Omega_{ip}$ can be written

\begin{equation}
    e^{-i \gamma H_P} e^{i \beta H_D} Z_{ip} e^{-i \beta H_D} e^{i \gamma H_P} = \exp( \frac{-i \gamma}{4} \Delta_{ip}) \exp(\frac{-i \gamma}{4} \Theta_{ip}) Y_{ip} \exp(\frac{i \gamma}{4} \Theta_{ip}) \exp( \frac{i \gamma}{4} \Delta_{ip}),
    \label{eq:y_conj}
\end{equation}
where 
\begin{equation}
\label{eq:Theta}
\begin{split}
    \Theta_{ip} &= \sum_{u,v}^N \mathbf{G_{uv}} \sum_{q=0}^k 2^q (Z_{qu} + Z_{qv}), \\
    \Delta_{ip} &= 2 \sum_{v=1}^N \sum_{s=0}^k \mathbf{G_{iv}} 2^{p+s} Z_{ip} Z_{vs}.
\end{split}
\end{equation}
Removing parts of $\Theta_{ip}$ which commute through $Y_{ip}$ allows the replacement of $\Theta$ with $\Theta'$, which is
\begin{equation}
    \Theta'_{ip} = 2^{p+1} \sum_{u=1}^N \mathbf{G_{iu}} Z_{ip}.
\end{equation}
Substituting $\Theta'$ for $\Theta$ in Eq. \eqref{eq:y_conj} and applying the transformation $R_Z (\Theta') Y R_Z (\Theta')^\dagger$ ($R_Z$ denoting the Bloch sphere rotation induced by a $Z$ gate) gives
\begin{equation}
    e^{-i \gamma H_P} e^{i \beta H_D} Z_{ip} e^{-i \beta H_D} e^{i \gamma H_P} = \exp( \frac{-i \gamma}{4} \Delta_{ip}) \Big[ \mathcal{C}_{ip} Y_{ip} -\mathcal{S}_{ip} X_{ip} \Big] \exp( \frac{i \gamma}{4} \Delta_{ip}).
    \label{eq:single_single}
\end{equation}
The last transformation to perform is conjugation of Eq. \eqref{eq:single_single} with the two-qubit gates of $H_P$.
When conjugated with $\ket{\psi_0}$, which is a Pauli X eigenstate, only tensor products of $X$ operators will give nonzero results. As such, one can see that the $Y_{ip}$ term of Eq. \eqref{eq:single_single} will disappear.
The remaining nonzero part of Eq. \eqref{eq:single} is 
\begin{equation}
\label{eq:single-double1}
    \Omega_{ip} = -\mathcal{S}_{ip} \bra{\psi_0} \exp( \frac{-i \gamma}{4} \Delta_{ip}) X_{ip} \exp( \frac{i \gamma}{4} \Delta_{ip}) \ket{\psi_0},
\end{equation}
Using the fact that $\bra{\psi_0} \otimes_t X_t \ket{\psi_0} = 1$, $\Omega_{ip}$ evaluates to the expression given in Eq. \eqref{eq:omega_def},
where the denominator derives from the fact that there is no $Z_{ip}Z_{ip}$ operator (this is the identity) and so should not be included in the product term.

\subsubsection{Two-qubit operator expectation}

The expectation of two-qubit operators $\Gamma_{ijpq}$ is the expectation of the two-qubit $Z_{ip}Z_{jq}$ interaction operator with respect to the final state

\begin{equation}
\begin{split}
\label{eq:gamma_def}
    \Gamma_{ijpq} &= \zeta_{ijpq}
    \Bigg[ {\cal S}_{ip}{\cal S}_{jq} \Bigg( \sum_{m=2x}^{Nk-2} \sum_{g}^{{Nk-2 \choose m}} \chi_{S_g} \Bigg) +
    {\cal C}_{ip}{\cal C}_{jq} \Bigg( \sum_{m=2x+1}^{Nk-2} \sum_{g}^{{Nk-2 \choose m}} \chi_{S_g} \Bigg) \Bigg], \\
    \zeta_{ijpq} &= \frac{\prod_{v=1}^N \prod_{r=0}^k \cos(2^{p+r} \gamma \mathbf{G_{iv}}) \cos(2^{q+r} \gamma \mathbf{G_{vj}})}{\cos(2^{2p} \gamma \mathbf{G_{ii}}) \cos(2^{2q} \gamma \mathbf{G_{jj}}) \cos^2(2^{p+q} \gamma \mathbf{G_{ij}})}.
    \end{split}
\end{equation}

where the function $\chi$ will be discussed in the following, and depends on combinatorial arguments.
The indices $(i,j,p,q)$ specify a qubit pair, one qubit indexed by $(i,p)$ and another by $(j,q)$. In the construction of the QAOA for SVP there are $Nk$ qubits in total, which comprise $N$ qudits, each comprising $k$ qubits.
In the formula for $\Gamma$ in Eq. \eqref{eq:gamma_def}, $S$ represents a subset of $m$ qubits chosen without replacement from the $Nk-2$ qubits remaining (i.e. not including qubits $(i,p), (j,q)$, of which there are ${Nk-2 \choose m}$ such subsets, distinguished by the subscript $g$ in $S_g$). Denote the qubits of a set $S$ of order $m$ to be 
\begin{equation}
\label{eq:qubit_subset}
S=\{ (t_1, w_1), \hdots, (t_m, w_m) \}.
\end{equation}
Then $\chi_S$ is given by Eq. \eqref{eq:chi} and is explained in detail in the following.
In the formula for $\Gamma$ in Eq. \eqref{eq:gamma_def}, one can treat the summations over $\chi$ in two parts. The first part, with prefactor ${\cal S}_{ip}{\cal S}_{jq}$ is a summation of $\chi_S$ for all (nonempty) subsets $S$ of qubits \textit{other than $(i,p), (j,q)$} which contain an even number of qubits. The second part, with prefactor ${\cal C}_{ip}{\cal C}_{jq}$ is a summation of $\chi$ for all subsets $S$ of qubits (also not including $(i,p), (j,q)$) with an odd number of qubits. For a given value of $m$, there are ${Nk-2 \choose m}$ distinct subsets of $m$ qubits, and these are indexed by the subscript $g$. Considering both parts together ($m$ odd and $m$ even), there are a total of $2^{Nk-2}-1$ terms. This exponential increase in computation is principally responsible for the difficulty in classical evaluation of $\mu$, and even the approximators $\mu_A$.

In order to derive the expression for $\Gamma$ of Eq. \eqref{eq:gamma_def} evaluate the expectation of $Z_{ip}Z_{jq}$ with respect to the evolved state of Eq. \eqref{eq:state} (with $-\gamma$ instead of $\gamma$)
\begin{equation}
\label{eq:double}
    \Gamma_{ijpq} = \bra{\psi_0} e^{-i \gamma H_P} e^{i \beta H_D} Z_{ip}Z_{jq} e^{-i \beta H_D} e^{i \gamma H_P} \ket{\psi_0}.
\end{equation}
The driver Hamiltonian, with $\beta= \pi/4$ has the effect of transforming $Z_{ip}Z_{jq}$ into $Y_{ip}Y_{jq}$.
Next, conjugate the $Y_{ip}Y_{jq}$ operators with the single-qubit part of $H_P$, given by $\Theta$ in Eq. \eqref{eq:Theta}. This means that the operators inside the expectation of Eq. \eqref{eq:gamma_def} are
\begin{multline}
    e^{-i \gamma H_P} e^{i \beta H_D} Z_{ip}Z_{jq} e^{-i \beta H_D} e^{i \gamma H_P} = \exp(\frac{-i \gamma}{4} (\Delta_{jq} + \Delta_{ip})) \exp(\frac{-i \gamma}{4} (\Theta_{jq} + \Theta_{ip})) Y_{ip} Y_{jq} \\ \exp(\frac{i \gamma}{4} (\Theta_{ip} + \Theta_{jq})) \exp(\frac{i \gamma}{4} (\Delta_{ip} + \Delta_{jq})).
    \label{eq:yy_conj}
\end{multline}
After applying the Bloch sphere rotations induced by $\Theta$, this is 
\begin{multline}
\label{eq:double-single2}
e^{-i \gamma H_P} e^{i \beta H_D} Z_{ip}Z_{jq} e^{-i \beta H_D} e^{i \gamma H_P} = \exp(\frac{-i \gamma}{4} (\Delta_{jq} + \Delta_{ip}))
    \Bigg[ \mathcal{C}_{ip} \mathcal{C}_{jq}  Y_{ip} Y_{jq} \\ - \mathcal{S}_{ip} \mathcal{C}_{jq} X_{ip} Y_{jq} - \mathcal{C}_{ip} \mathcal{S}_{jq} Y_{ip} X_{jq} + 
    \mathcal{S}_{ip} \mathcal{S}_{jq} X_{ip} X_{jq} \Bigg] \exp(\frac{i \gamma}{4} (\Delta_{jq} + \Delta_{ip})).
\end{multline}
The only terms of interest in Eq. \eqref{eq:double-single2} are those that do not go to zero after conjugation by the two-qubit part of $H_P$, induced by $\delta$, when measured in the $H_P$ basis. Because $\ket{\psi_0}$ is an $X_l$ eigenstate with eigenvalue $1$, and $\bra{\psi_0} Y_l \ket{\psi_0}, \bra{\psi_0} Z_l \ket{\psi_0} = 0$ for all $l$, the only terms to survive are products of $X_l$'s with no trailing $Y_l, Z_l$.

In the following we consider how two-qubit $ZZ$ gates flip $Y$ operators to $X$ and back again. As all-$X$ tensor products are required, each operator must therefore have been flipped the right number of times modulo $2$, and there must be no trailing $Z, Y$ operators in the final term. Applying all two-qubit $ZZ$ operators to each of the terms in Eq. \eqref{eq:double-single2} gives the following results case by case, where a single index in ($a,b,c,d$) is used to distinguish qubits:

\textit{Case 1 - $X_a X_b$} Only $XX$ terms will give a nonzero expectation when conjugated with the initial state $\ket{\psi_0}$, so after applying all $ZZ$ gates in $H_P$, one should only consider the $XX$ operators with no trailing $Z,Y$. 
After applying all $ZZ$ gates, there will be one $XX$ term with a coefficient which is a product of all $\cos$'s.
Applying $Z_a Z_c$ and $Z_b Z_c$ results in a $Y_aY_b$ term, and then applying $Z_aZ_d$ and $Z_bZ_d$  gives another $X_aX_b$ term with no trailing $Y,Z$ operators, and a coefficient which is a product of $\sin$'s. After acting all other two-qubit gates on this $XX$ operator, a nonzero term consisting of four $\sin$'s and $2(Nk-4)$ $\cos$'s is obtained. Qubits $c,d$ were used to flip the $X_a X_b$ to $Y_a Y_b$ and back to $X_a X_b$, and there are ${Nk-2 \choose 2}$ ways to pick qubits $c,d$. This can be repeated iteratively, as the same result holds when using qubits $c,d,e,f$ to perform the operator flipping, and so there is a corresponding $X_a X_b$ term (with no trailing $Y, Z$) for each distinct combination of $\omega$ qubits other than $a,b$, where $\omega$ is even. This means there are ${Nk-2 \choose \omega}$ $XX$ terms for each even $\omega \leq Nk-2$.

\textit{Case 2,3 - $X_a Y_b, Y_a X_b$} These yield no nonzero terms. This is because only one of $X_a, Y_b$ needs to be flipped, but this results in a trailing $Z$ meaning that no $X_a X_b$ terms with no trailing $Y, Z$ are present in the final expression.

\textit{Case 4 - $Y_a Y_b$} This follows the same reasoning as for {\it Case 1}. The difference is that instead of requiring even numbers of qubits other than $a,b$, an odd number is required, and so for $\omega=1,3,\hdots$  there are ${Nk-2} \choose {\omega}$ nonzero terms, each of which are composed of a product of $2(Nk-2) - 2\omega$ $\cos$ terms and $2\omega$ $\sin$ terms.

Using this rationale, as a demonstrative example, let us deduce the expectation of a single $XX$ operator transformed by the action of the two-qubit part of $H_P$. For reference, this is equivalent to evaluating the final term in the central square brackets of Eq. \eqref{eq:double-single2}, without the trigonometric prefactors $\mathcal{S}_{ip}, \mathcal{S}_{jq}$.
The system is fully connected, and the only two-qubit gates to give special consideration to are $Z_{ip}Z_{ip}, Z_{jq}Z_{jq}$ which are the identity, and $Z_{ip}Z_{jq}$, which commutes through $X_{ip}X_{jq}$. In total there are $2^{Nk - 2}$ nonzero terms, and the term where the $X_{ip}X_{jq}$ is not flipped at all has the value $\zeta_{ijpq}$ from Eq. \eqref{eq:gamma_def},
where the denominator has the effect of cancelling the gates $Z_{ip}Z_{ip}, Z_{jq}Z_{jq}, Z_{ip}Z_{jq}$, from the product term. The value $\zeta$ is used as a stem function for all the remaining nonzero terms, which are a product over the same indices, and with the same angles, but with $\sin$'s in lieu of $\cos$'s. 

Now update $\zeta$ for a given combination of qubits which perform the operator flips (from $XX$ to $YY$ and back). 
Then the nonzero part of the expectation deriving from this combination of operator flips is $\zeta_{ijpq} \chi_S$, where $S$ is given in Eq. \eqref{eq:qubit_subset} and $\chi_S$ is 
\begin{equation}
\label{eq:chi}
    \chi_S = \prod_{h=1}^d \tan(2^{p+w_h} \gamma \mathbf{G_{i t_h}}) \tan(2^{q+w_h} \gamma \mathbf{G_{t_h j}}).
\end{equation}
Considering all combinations of qubits, and incorporating into \textit{Cases 1, 4} for Eq. \eqref{eq:double-single2}, one arrives at the value for $\Gamma$ given in Eq. \eqref{eq:gamma_def}.

\bibliography{references}

\end{document}

\beginsupplement

\title{Supplementary Material}

\author{David Joseph}
\affiliation{Electrical and Electronic Engineering Department, Imperial College London}%
\affiliation{Physics Department, Imperial College London}

\author{Antonio Martinez}
\affiliation{Sandbox@Alphabet}%

 \author{Cong Ling}
 \affiliation{Electrical and Electronic Engineering Department, Imperial College London}%
 
 \author{Florian Mintert}
 \affiliation{Physics Department, Imperial College London}%

% \begin{abstract}
% What's in the Supp. Mat.
% \end{abstract}

% \tableofcontents

\date{\today}%10 Feb 2021}

\maketitle

\bibliographystyle{plainnat}
\bibliography{references}

\makeatletter\@input{xx.tex}\makeatother